\documentclass{article}
\usepackage[ansinew]{inputenc}
\usepackage[centertags,reqno]{amsmath}
\usepackage{epsfig}
\usepackage{multirow}
\usepackage{graphicx}
\usepackage{rotating}
\usepackage[left=3cm,right=3cm,top=3cm,bottom=3cm]{geometry}
\usepackage{algorithmic}
\usepackage{algorithm}
\usepackage{clrscode3e}%algorithms

\title{Accelerating unstructured finite volume computations on field-programmable gate arrays}
\author{Zoltán Nagy
\thanks{Cellular Sensory and Wave Computing Laboratory, Computer and Automation Research Institute, Hungarian Academy of Sciences, 1518 Budapest, Pf. 63., Hungary, Email: nagyz@sztaki.hu}
\and Csaba Nemes
\thanks{Faculty of Information Technology, Péter Pázmány Catholic University
}
\and Antal Hiba$^{\dag}$
\and Árpád Csík
\thanks{Széchenyi István University, Department of Mathematics and Computational Sciences, Gy\H{o}r, Hungary}
\and András Kiss$^{* \dag}$
\and Miklós Ruszinkó
\thanks{Applied Mathematics Research Laboratory, Computer and Automation Research Institute, Hungarian Academy of Sciences, Budapest, Hungary}
\and Péter Szolgay$^{* \dag}$
}
\date{}

\begin{document}
\maketitle
\begin{abstract}
Accurate simulations of various physical processes on digital computers requires huge computing performance, therefore accelerating these scientific and engineering applications has a great importance. Density of programmable logic devices doubles in every 18 months according to Moore's Law. On the recent devices around one hundred double precision floating-point adders and multipliers can be implemented. In the paper an FPGA based framework is described to efficiently utilize this huge computing power to accelerate simulation of complex physical spatio-temporal phenomena. Simulating complicated geometries requires unstructured spatial discretization which results in irregular memory access patterns severely limiting computing performance. Data locality is improved by mesh node renumbering technique which results in predictable memory access pattern. Additionally storing a small window of node data in the on-chip memory of the FPGA can increase data reuse and decrease memory bandwidth requirements. Generation of the floating-point data path and control structure of the arithmetic unit containing dozens of operators is a very challenging task when the goal is high operating frequency. Long and high fanout control lines and improper placement can severely affect computing performance. In the paper an automatic data path generation and partitioning algorithm is presented to eliminate long delays and aid placement of the circuit. Efficiency and use of the framework is described by a case study solving the Euler equations on an unstructured mesh using finite volume technique. On the currently available largest FPGA the generated architecture contains three processing elements working in parallel providing 90 times speedup compared to a high performance microprocessor core.
\end{abstract}

\section{Introduction}
Numerical simulation of complex problems evolving in time plays an important role in scientific and engineering applications. Accurate behavior of dynamical systems can be understood using large scale simulations which traditionally requires expensive supercomputing facilities.  Several previous studies proved the efficiency of programmable logic devices in numerical simulation of various physical phenomena such as electromagnetic \cite{Durbano2004}, transient wave \cite{He2005} \cite{Sonkoly2007} and computational fluid dynamics \cite{Sano2007} \cite{KOCSARDI2008} simulations.

A complex spatio-temporal problem can be approximated in a regular mesh structure. To get a more accurate solution the the resolution of the mesh should be increased, which increases the number of grid points and the running time as well. To overcome the outlined problem an unstructured mesh will be used. The resolution of the mesh will be increased where it is required by rapid change in the dynamics or shape of the problem. Conventional microprocessors has around $10\%$ utilization during unstructured mesh computations due to the irregular memory access pattern of grid data.

To accelerate the computation irregular memory access patterns should be hidden by temporally storing the relevant grid points on the FPGA. Due to the finite on-chip memory size mesh points have to be ordered and stored according to the requirements of computation. A very efficient new mesh point ordering method was developed.

%As it has been shown to use a higher operating frequency – increasing the operating speed as a consequence of this - is limited due to the wire delays of the global control signals.
To obtain a high performance accelerator the operating frequency of the arithmetic unit is critical.
By using local controls to clusters of  arithmetic  operators the clock speed can be significantly increased paying with some area increase for it. The data transfer between the clusters are synchronized by FIFOs. It was proofed that a good tradeoff can be achieved if a cluster has about 10 I/O FIFOs.  To define the clusters several partitioning algorithms \cite{Nemes2010} \cite{Nemes2011} were carefully tested and compared. The placement and the routing of the clusters however were not as good as it should be. Especially due to the long interconnections the clock frequency could not increased further. To eliminate this kind of problem a new combined  the clustering and placement  method will be shown and tested in a complex fluid dynamical problem. 

%The mesh is composed of non-overlapping triangular elements, called finite volumes. The flow variables are stored at the mass center of the triangles. The governing equations are discretized by an explicit finite volume method employing the simple Lax-Friedrich numerical flux function.
In Section 2. an overview of recent publications on accelerators working on unstructured meshes are given. A new architecture for efficient unstructured mesh computations is proposed in Section 3. A new node reordering algorithm is presented in Section 4. and a new partitioning algorithm is descried in Section 5. The proposed methods are tested on a complex case study described in Section 6. Finally performance of the proposed architecture is compared to a high performance microprocessor in Section 7.

\section{Related work}
Several papers are published in the early 2000s dealing with the acceleration of PDEs on unstructured meshes. Most of them are focused on accelerating Finite Element Methods (FEM) where the global stiffness matrix is given and the resulting large linear system of equations are solved usually by the iterative Conjugate Gradient (CG) method. The most time consuming operation of this method is a sparse matrix vector multiplication therefore most of the papers try to accelerate this part. Though our architecture is designed for explicit unstructured finite volume calculations examination of these architectures is helpful because similar problems arising in our case such as the adjacency matrix of the mesh is sparse and elements of the state vector are accessed in a random order.

In 2000 Jones and Ramachandran \cite{Jones2000} examined several aspects of accelerating unstructured mesh computations on FPGAs. They proposed a hybrid architecture to accelerate the CG algorithm where the local stiffness matrix and the bulk of the CG algorithm is computed by the CPU and only the matrix vector multiplication is performed on the FPGA and achieved $22.4-35.7$MFLOPs computing performance. 

Another approach is the architecture proposed by deLorimier and DeHon \cite{deLorimier2005} where all elements of the matrix are stored on the FPGA to avoid bandwidth limitations but this solution severely limited the size of the matrix. Performance of the architecture is depended on the structure of the matrix and usually $66\%$ of the peak performance can be achieved which results in 2-10 times acceleration compared to the common microprocessors at that time.

Elkurdi et al. \cite{Elkurdi2008} first reorganize the finite element sparse matrix into a banded form then calculate the matrix vector multiplication along special pipelineable diagonal stripes, where two successive elements can be processed by a pipelined architecture. Performance of the architecture is determined by the available memory bandwidth and the sparsity of the matrix however utilization of the processing elements is varying in a very wide range between $17.74-86.24\%$. 

duBois et al. \cite{dubois2010} presented an architecture where nonzero elements from each row of the sparse matrix are processed in 7 element wide vectors. They also proposed to use a reordered banded matrix to improve data locality, but the architecture still suffer from memory bandwidth limitation.

Recently Nagar et al. \cite{Nagar2011} proposed an architecture using an optimized Streaming Multiply- Accumulator with separate cache memories for matrix and vector data. The implementation platform they used has special memory architecture providing high 20GB/s peak memory bandwidth. Performance of the system with four FPGAs is in the $1.17-3.94$GFLOPs range outperforming a Tesla 1070 GPU. However utilization of the PEs is around $50\%$ similarly to the previous architectures and increasing the number of PEs to occupy the entire FPGA still runs into a memory bandwidth limit.

The surveyed architectures provide general solutions to accelerate FEM calculations on FPGAs but suffer from the inherent high memory bandwidth requirement and small communication to computation ratio of sparse matrix vector multiplication. On the other hand utilization of the execution units depends on the structure of the sparse matrix. 

In the case of finite volume discretization irregular memory access pattern and high memory bandwidth requirement can be eliminated by storing a small part of the grid on-chip and reordering its nodes. Right hand side of the discretized equations should be computed for each node in every timestep, which requires several floating-point operations, resulting in better communication to computation ratio and higher utilization of FPGA resources. 
%In case of linear PDEs performance can be increased by using matrix-free implementation where elements of the local stiffness matrix is not stored but computed on the FPGA in every iteration.

\section{Architecture}
\begin{figure}
\begin{center}
\epsfig{width=14cm,figure=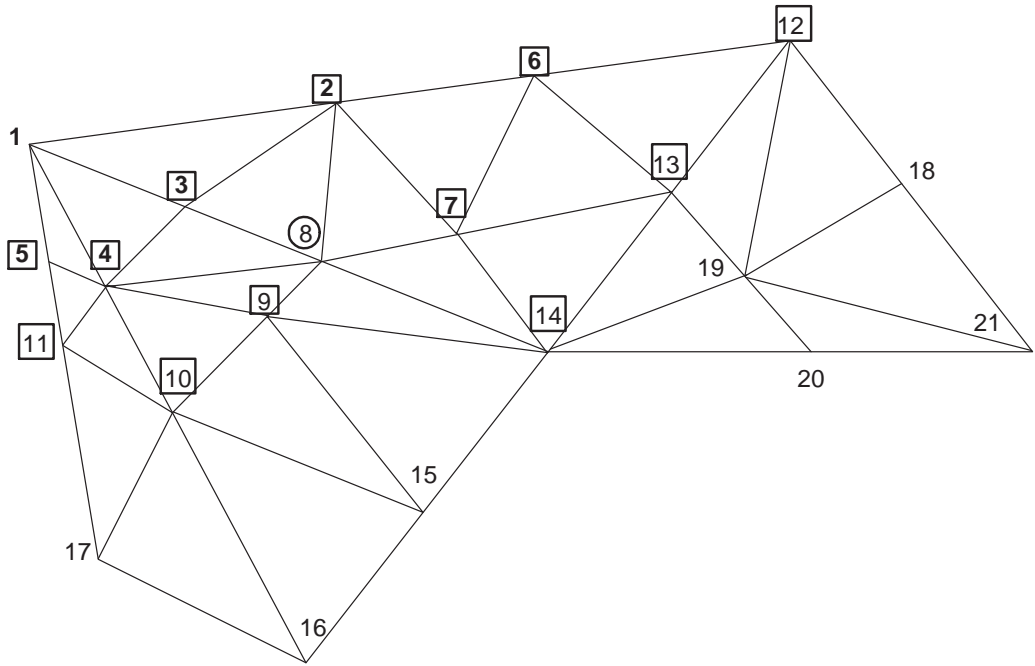}
\end{center}
\caption{Local node data stored on the FPGA} \label{FigMemContents}
\end{figure}

Time evolution of dynamical systems can be easily solved numerically by explicit finite volume discretization and the solution is computed on a discrete set of points in space. The solution is advanced in time by approximating the derivative of each node using linear combination of state values from its small local neighborhood. Computation of subsequent grid points is independent however some parts of its input data sets are overlapped. For practical applications the number of grid points is far exceeding the available on-chip memory of recent FPGAs therefore state and constant values must be stored and loaded from an off-chip memory. Overlapping input data sets can be utilized to reduce the number of memory accesses and save memory bandwidth by saving all nodes which required during the following computations into a local memory on the FPGA. Simple example for the values to be stored is shown in Figure~\ref{FigMemContents}.

Each node is numbered and the nodes are updated in an increasing order, bold numbers indicate the already processed nodes, encircled node 8 is the currently processed vertex, while squared nodes are stored on the FPGA. In this case all elements required for the computation of node 8 can be read from the fast on-chip memory of the FPGA. To process the next node a new node 15 should be loaded and node 2 can be discarded. When updating node 9 all the required data will be available in the on-chip memory. It is possible that multiple new nodes are required for the update of a node indicating that the on-chip memory is undersized. The size of the required on-chip memory depends on the structure of the grid and the numbering of the nodes. 

\begin{figure}
\begin{center}
\epsfig{width=14cm,figure=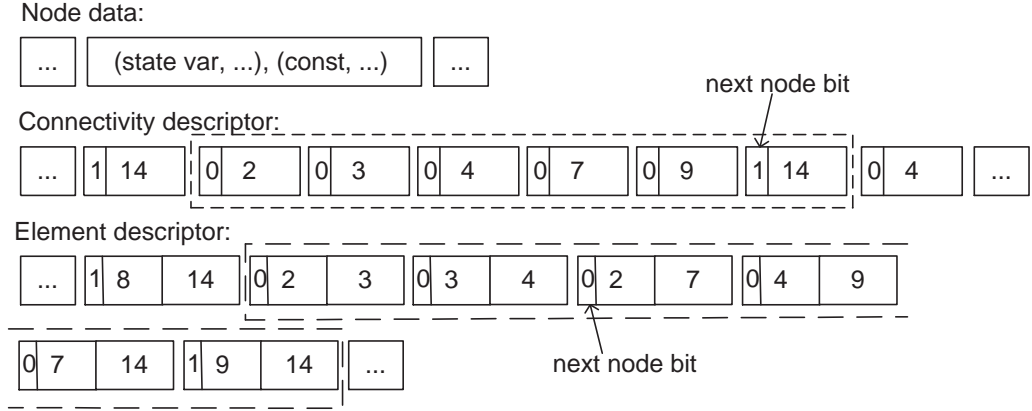}
\end{center}
\caption{Data structures} \label{FigDataStruct}
\end{figure}

Node data is treated as a 1D array where each element contains two separate parts one for storing time dependent state values and one to store constant data such as physical constants and coordinates of the grid points. In case of structured grids neighbors of each grid point can be determined by its array index. For unstructured grids connected grid points are described by a sparse adjacency matrix which is usually stored in a Compressed Row Storage (CRS) format \cite{Duff1989}. In our case the matrix contains only 0 and 1 elements therefore only column indices are stored. Additionally the elements are read in a serial sequence hence row pointers can be replaced by a single bit to indicate the start of a new row. It is possible that the discretization stencil is defined on triangle or tetrahedron instead of a line between two grid points. In this case an additional element descriptor is required where node indices of the vertexes of the element are stored. Data structures used by the system are shown in Figure~\ref{FigDataStruct}. Additionally an example data structure is filled with connectivity and element data for node 8 from Figure~\ref{FigMemContents}. Width of the index field can be set according to the requirements of the application. More than 8 million nodes can be handled by using 24 bit wide indices which can be sufficient in many applications.  

\begin{figure}
\begin{center}
\epsfig{width=14cm,figure=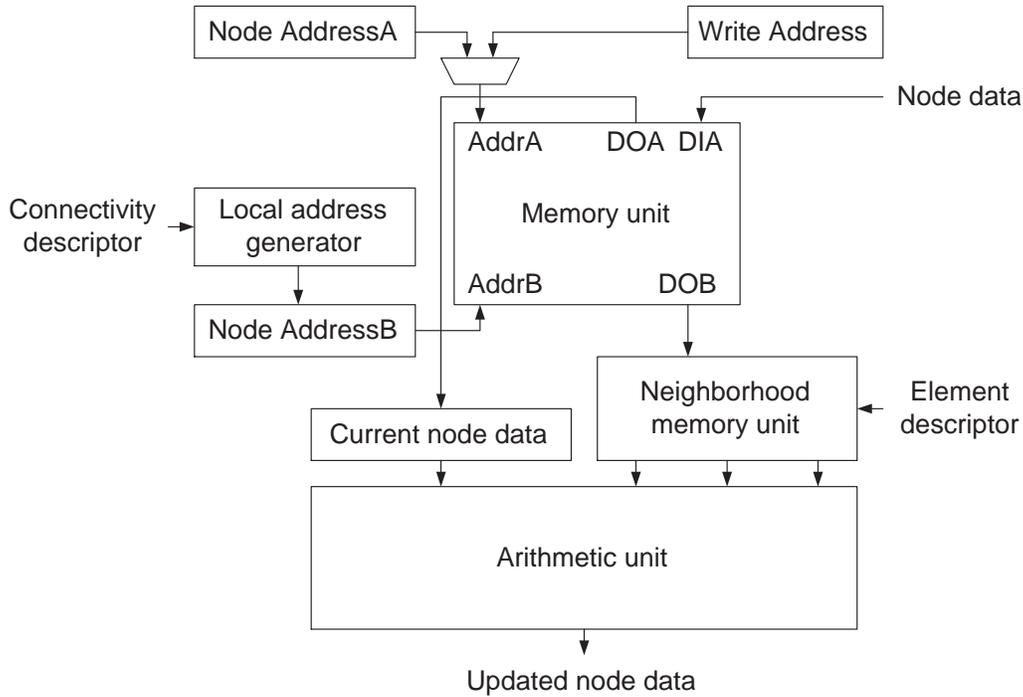}
\end{center}
\caption{Block diagram of the proposed architecture} \label{FigUnstructArch}
\end{figure}

Main parts of the proposed architecture are the Memory unit, the Neighborhood memory unit and the Arithmetic unit as shown in Figure~\ref{FigUnstructArch}. The Memory unit is build from dual ported on-chip BRAM memories and store a small part of grid data.
%Size of the on-chip memory should be at least twice the bandwidth of the adjacency matrix of the mesh.
Size of the on-chip memory is determined by the bandwidth of the adjacency matrix of the mesh. Bandwidth of a matrix is defined as the maximum distance of nonzero element from the main diagonal. (See Section~\ref{ParSerialBW} for a more formal definition.)
Neighbors of the currently processed nodes are stored in the Neighborhood memory unit. Its structure is depending on the particular discretization stencil. In the simplest case when the stencil is defined on a line it can be a simple register. When the stencil is defined on a triangle or tetrahedron a one write two or three read multi ported memory is required respectively. Size of this memory is depending on the largest node rank in the mesh, usually a 64 element memory is sufficient, which can be efficiently implemented using the Distributed RAMs on FPGAs. Computation of the updated node value is carried out in an element by element order by the arithmetic unit. 

Computation is started by loading node values into the Memory unit until it is half filled. In this phase value of the first node is loaded to the Current node register and the Neighborhood memory is filled by its neighbors using the incoming connectivity descriptors. Global indices of the neighboring nodes are translated into addresses in the Memory unit by the Local address generator unit. When all neighbors are loaded valid stencil data can be send to the arithmetic unit in each clock cycle. Neighborhood of the second node can be loaded into the free space of the Neighborhood memory while processing the first node. When all parts of the first stencil are sent to the arithmetic unit Node addressA register is incremented and the second node is loaded into the Current node register. During the next clock cycle new node data can be written into the Memory unit and computation of the second node is started along with the loading of the neighborhood of the third node. The Memory unit is operating as a circular buffer when it is filled the oldest node data is overwritten which can be done safely because the size of the memory is set to be at least twice the bandwidth of the adjacency matrix and one can be sure that these old values are never required during the update of the remaining nodes.

Connectivity and element descriptors add an overhead for the off-chip memory requirements and increase memory bandwidth requirement of the processor. The width of the index fields can be optimized according to the depth of the Memory and Neighborhood memory units because all memory accesses are hidden by these two units. Therefore the global index of the nodes are never required only the translated memory addresses for the on-chip memories which can be much shorter. The order of node updates are statically scheduled hence memory address translation can be done offline by the host CPU. 

The main advantage of the proposed architecture is the serial off-chip memory access pattern on the node data and descriptor arrays. Each array is read into a FIFO buffer by using an optimal burst length and penalties of random memory access patterns can be eliminated. Maximum size of the mesh is limited by the bandwidth of the adjacency matrix. Using today large, high performance FPGAs even 10,000-40,000 nodes can be stored on-chip and depending on the structure of the mesh its size can be in the 100,000-400,000 node range.

\section{Memory Bandwidth Optimization}
Efficient use of on-chip memory resources of the architecture described in the previous section is depending on the bandwidth of the adjacency matrix of the mesh. By reordering the nodes of the mesh the bandwidth of the adjacency matrix can be significantly reduced. In this section we show a fast constructive method for reordering, which is comparable to the classical algorithms, and we present a solution for generating memory access patterns which have lower bandwidth than a given bound.

\subsection{Description of the problems and related works}
\paragraph{Matrix Bandwidth Minimization:}
Let G(V,E) a graph with vertex set V, \(|V|=n\) and edge set E. Labeling is a function f(v) which assigns integers[1..n] to vertices. f(v)=f(u) if and only if u=v \(u,v\in V\). N(v) is the set of vertices which adjacent to v. The bandwidth of a vertex v is \(B_f(v)=Max\{|f(v)-f(u)|: u \in N(v)\}\), and the bandwidth corresponds to G with labeling f() is \(B_f(G)=Max\{B_f(v): v\in G\}\).\\
The problem of bandwidth reduction of a sparse matrix was shown to be NP-complete by Papadimitriou\cite{Papadimitriou}, so exact methods can not be applied for large problems. Many heuristic algorithm was developed from the well-known Cuthill-McKee (CM) algorithm\cite{CM-origin} to recent metaheuristic approaches. One of the most promising metaheuristic results in solution quality is the GRASP (Greedy Randomized Adaptive Search Procedure) with Path Relinking, which was presented in\cite{GRASP-PR}. The most practical solution is the GPS(Gibbs, Poole and Stockmeyer) method\cite{GPS-origin}, which is one of the fastest heuristics, and also provides good solution quality. Metaheuristic methods gives better solutions, but their time consumption many times higher than GPS. GPS algorithm was born in 1976, afterwards many attempts shown to improve the original method, the most interesting from them is the method created by Luo\cite{GPS-Luo}. All of the improved GPS heuristics have higher time complexity, which is an important parameter in case of large meshes.

\paragraph{Serial-Bandwidth:}\label{ParSerialBW}
Using the classical definition of bandwidth, the size of the on-chip memory can be given as \((B_f(G)*2 + 1)*sizeof(node)\), where \((B_f(G)*2 + 1)\) called C\_BW. If we assume the architecture described in previous section, a more proper definition than C\_BW can be given. S\_BW means the number of nodes which have to be stored in on-chip memory, if we want 0\% cache-miss.
\[s(i)=MIN\{f(v):v\in N(u), f(u)=i\}\] 
\[e(i)=MAX\{f(v):v\in N(u), f(u)=i\}\]
\[S(i)=MIN\{s(i), s(i+1), ... ,s(n)\}\]
\[E(i)=MAX\{e(1), e(2), ... ,e(i)\}\]
\[S\_BW = MAX_{i}\{ E(i)-S(i) \}\]

\subsection{Algorithm for bandwidth reduction}
Several methods have been shown in the literature for minimizing the classical bandwidth of a system. In this section we define Amoeba1(AM1) algorithm for serial-bandwidth minimization. Our goal is to create a fast, effective constructive method which has proper, easy to calculate S\_BW bounds in each construction step(details in next subsection).

\subsubsection{Notations and definitions}
Amoeba1 is a constructive method, in which a solution element is chosen and labeled in each step. Solution elements are the vertices of the input mesh, and the method grows a part till all of the vertices are covered. \\
\begin{figure}[h!]
  	\centering
	\epsfig{width=0.5\textwidth,figure=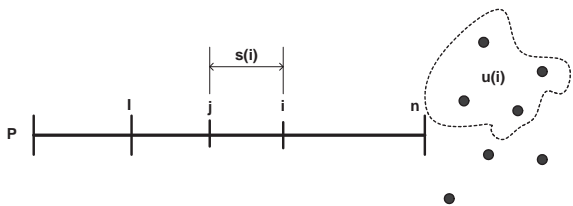}
    \caption{Structure of solution part P.}
    \label{fig:part}
\end{figure}

Figure~\ref{fig:part}. shows the structure of a solution part P with n elements. Every node(i) have three base parameters: local\_index=i, s(i), u(i).
\begin{description}
\item{\textbf{s(i):}}  is the distance between node(i) and its  lowest indexed neighbor in the part: \(s(i)=MAX\{i-j : j \in N(i)\}\).
\item{\textbf{u(i):}} is the set of nodes which uncovered by P, but must be added in later steps because of node(i): \(u(i)=\{v : v \in N(i) \; AND \; v \notin P\}\).
\item{\textbf{I:}} \quad is the index of the first elemet which has not empty u() set, so for every node(i) where $i<I$ all neighbors covered by P.
\end{description}
With these parameters we can give bounds on the serial bandwidth. In AM1 method we use a simple lower bound for describing the importance of node(i):

\[imp(i)=(n-i)+|u(i)|+s(i)\]

This is obviously a lower bound, because if we add node \(v\notin u(i)\) to part P we still have to add all elements of u(i) to the part. For every node(i) $i<I$ imp(i)=0, because these nodes have all of their neighbors involved, so their effect on bandwidth do not depend on the later decisions. \\

\subsubsection{Description of Amoeba1}
Amoeba1 algorithm has two base steps: finding a starting vertex, and the labeling loop. The result is an ordering of the vertices.
\paragraph{Finding a starting vertex}
The quality of the result of constructive bandwidth-reduction heuristics depends on which is the starting vertex. In GPS method, the authors presented a simple and effective solution for this problem. They gave an algorithm which returns the two endpoints of a pseudo-diameter. AM1 algorithm uses this subroutine for finding the starting vertex.

\paragraph{Choosing a solution element}
AM1 selects a node from u(I), which has a neighbor in P with maximal importance. Because all nodes in u(I) has node(I) as its neighbor, only $l \neq node(I)$ neighbors take part in the search.
\begin{codebox}
\zi \Comment selecting candidate from u(I)
\li $candidate = random \; element \; of \; u(I)$
\li $global\_max=0$
\li \For $\forall k \in u(I)$
\li \Do
			$local\_max=0$
\li		\For $\forall l \in N(k): \; l\in P \; and \; l\neq node(I)$
\li		\Do
				\If $l.imp()>local\_max$
\li			\Then $local\_max=l.imp()$
				\End
			\End
\li	\If $local\_max > global\_max$
\li	\Then
			$candidate = k$
\li		$global\_max=local\_max$
		\End
		\End
\end{codebox}

AM1 adds the candidate to the part with index=n+1, and chooses the next element till the whole mesh is indexed. AM1 performs a kind of breadth-first indexing.

\subsubsection{Results and conclusions}
AM1 is a simple constructive algorithm for large problems, we compare its results to the classical fast and effective GPS method. As mentioned earlier, better quality algorithms exist for bandwidth reduction, but these methods can not be applied to large meshes(>100.000 vertex) because of their complexity. The cases showed on Table~\ref{AM1-2d} comes from 2-dimmensional meshes, with assigning a vertex to each triangle and an edge between vertices which are represent adjacent triangles, so we get a mesh with maximal\_degree = 3. This meshes appears when we use finite volume solver during the solution of a partial differential equation. In these low-degree cases AM1 provides similar solution quality to GPS, in 4\% less time. The running time of both method depends on the number of vertices, and the structure of the mesh(finding a starting vertex).

\begin{table}[h!]
\begin{center}
\caption{Results of Amoeba1 method compared to GPS.}
\label{AM1-2d}
\setlength{\tabcolsep}{3pt}
\begin{scriptsize}
\begin{tabular}{|c|c|c|c|c|c|}
\hline
Case & N & S\_BW GPS & S\_BW AM1 & GPS time(s) & AM1 time(s)\\ \hline
step\_2d\_bc\_cl30 &	7063	& 122 &	122 &	0,078 &	0,052\\ \hline
step\_2d\_bc\_cl40 &	12297	& 176 &	175 &	0,154 &	0,109\\ \hline
step\_2d\_bc\_cl50 &	20807	& 253 &	227	& 0,175	& 0,1\\ \hline
%step\_2d\_bc\_cl60 &	31241	& 311	& 284	& 0,448	& 0,343\\ \hline
step\_2d\_bc\_cl70 &	42449	& 359	& 341	& 0,633	& 0,49\\ \hline
%step\_2d\_bc\_cl80 &	54839	& 421	& 402	& 0,779	& 0,596\\ \hline
step\_2d\_bc\_cl90 &	68271	& 481	& 506	& 0,998	& 0,785\\ \hline
%step\_2d\_bc\_cl100&	90839 &	528	& 487	& 1,716	& 1,458\\ \hline
step\_2d\_bc\_cl110&	112093&	569	& 591	& 2,144	& 1,955\\ \hline
%step\_2d\_bc\_cl120&	131465&	626	& 658	& 2,204	& 1,86\\ \hline
step\_2d\_bc\_cl130&	157099&	740	& 738	& 1,59	& 1,316\\ \hline
%step\_2d\_bc\_cl140&	179119&	751	& 790	& 2,828	& 2,525\\ \hline
step\_2d\_bc\_cl150&	201069&	794	& 805	& 3,239	& 3,094\\ \hline
%step\_2d\_bc\_cl160&	227139&	853	& 852	& 2,398	& 2,069\\ \hline
step\_2d\_bc\_cl170&	252869&	972	& 923	& 4,316	& 3,92\\ \hline
%step\_2d\_bc\_cl180&	280789&	982	& 963	& 6,924	& 6,547\\ \hline
step\_2d\_bc\_cl190&	316715&	1030&	1082&	5,913	& 5,707\\ \hline
step\_2d\_bc\_cl200&	394277&	1093&	1155&	5,855	& 5,532\\ \hline
step\_2d\_bc\_cl320&	930071&	1923&	1809&	17,035&	18,687\\ \hline
\multicolumn{6}{l}{S\_BW: serial-bandwidth of solutions, N: number of vertices.}\\
\multicolumn{6}{l}{Algorithms tested on one core of an Intel P8400 processor}\\
\end{tabular}
\end{scriptsize}
\end{center}
\end{table}

The results for high-degree(20-30) cases can be found on Table~\ref{AM1-3d}. These cases generated from the same complex 3D geometry, by increasing the density of the mesh. We found that GPS is 29\% superior on these general instances, but 13\% slower than AM1. These results shows, that the difference is not increasing with the complexity of the problems. In our further work, we want to increase solution quality by guiding the labeling depending on the knowledge of the geometry.

\begin{table}[h!]
\begin{center}
\caption{Results for 3D high-degree cases}
\label{AM1-3d}
\setlength{\tabcolsep}{3pt}
\begin{scriptsize}
\begin{tabular}{|c|c|c|c|c|c|}
\hline
Case & N & S\_BW GPS & S\_BW AM1 & GPS time(s) & AM1 time(s)\\ \hline
3d\_075 & 3652 &	381 &	391 &	0,279 &	0,27\\ \hline
3d\_065 & 5185 &	500 &	763 &	0,144 &	0,107\\ \hline
3d\_055 & 8668 &	712 &	764 &	0,655 &	0,587\\ \hline
3d\_045 & 15861&	1066&	1478&	0,468	& 0,42\\ \hline
3d\_035 & 33730&	1880&	1863&	2,209	& 2,22\\ \hline
3d\_025 & 88307&	3384&	3443&	7,569	& 6,42\\ \hline
3d\_018 & 244756 &	6582 &	10110	& 39,509	& 27,598\\ \hline
3d\_015 & 417573	& 9066 &	14930	& 85,797	& 59,958\\ \hline
3d\_012 & 519983	& 20561&	23554	& 413,72	& 383,075\\ \hline

\end{tabular}
\end{scriptsize}
\end{center}
\end{table}

\subsection{Memory Access Optimization for Bounded Bandwidth}
In case of large problems it is possible, that the renumbered mesh has larger bandwidth than the available on-chip memory, these cases should be handled too. In this section we show an AM1 based method, which generate an input order which has at most a pre-specified serial-bandwidth. In the input order every vertex executed once, but can be loaded many times, so we need a flag \textbf{Ex} to store it is only a ghost node(false) or have to be calculated(true). Serial bandwidth in this job means the following: for every Ex=true vertex all of their neighbors surely be in the on-chip memory when the execution reaches them. If the defined bound is less than the serial-bandwidth for whole mesh provided by the AM1 method, the input will be k-times longer, where \(1\leq k\). The bound on bandwidth obviously have to be more than the maximal\_degree of the graph.

\subsubsection{AM1 based bounded S\_BW method}
The main concept of handling the bounded bandwidth is the usage of a proper serial bandwidth estimation, which is available in AM1 method. When a Part reaches the S\_BW bound, the process calculates which vertices can be executed, and call the AM1 method for the rest of vertices where Ex is not true. The main process starts new AM1 parts until all vertices are Ex=true. The output of the method is an access pattern(list of \{index,Ex\} pairs), where all vertex has true execute flag once, and can be appear many times as a ghost node.
\paragraph{Estimation of serial-bandwith:}
Given an AM1 Part, the task is to estimate its serial bandwidth. AM1 estimates the part's bandwidth in each construction step. If $i<I$ for node(i) in part P, it has all of its neighbors inside the part, so if the bandwidth is less than the bound when I become larger than k, node(k) can not increases S\_BW anymore. As shown earlier imp(i) is a lower bound on serial-bandwidth, but in the estimation  a proper upper bound is required for the stopping condition. Because in all step AM1 adds a node from u(I) to the part, we can calculate more than a proper upper bound for node(i), we can give the exact value.
\begin{equation}\label{exact-sbw}
S\_BW(i) = (n-i) + |\bigcup_{I\leq k \leq i}u(k)| + s(i)\quad I\leq i
\end{equation}
Eq.~\ref{exact-sbw} is equivalent to the definition of serial bandwidth\footnote{the authors have a proof} Equation~\ref{exact-sbw} could be a stopping condition, but the proposed method has a less complex and useful upper bound for stopping decision defined in Eq.~\ref{condition}.
\begin{equation}\label{condition}
S\_BW Bound \geq \underbrace{MAX}_{I \leq k \leq n}imp(k)
\end{equation}
If Eq.~\ref{condition} holds, AM1 continue to add nodes to part P, stop otherwise. This condition is not an upper bound for the whole part, but it provides in every step that node(I) has lower serial-bandwidth than the given bound. If I jumps to I' when AM1 adds node(n) to P, we can be sure that for all node(i) \(I\leq i \leq I'\) serial bandwidth is under the bound, because \(\bigcup_{I\leq k \leq I'}u(k) = \{node(n)\}\), so \(imp(i)=S\_BW(i)\) inside the range [I ,I'].
\paragraph{Finalizing a Part:}
When Eq.~\ref{condition} not holds, the proposed algorithm finalize the part, and starts a new instance of AM1 on the rest of the not executed nodes. Finalization have two tasks: it has to label vertices which executed in part P, and have to label vertices which all neighbors executed too, because these nodes can be cut out of the mesh(we call them perfect nodes Pr=true). Ex=true and Pr=false vertices have to be loaded again, because they have at least one Ex=false neighbor. In AM1 imp(i)=0 and u(i)=\{\} for all node(i) for which Ex=true. 
\[s^* = \underbrace{MIN}_{I\leq k \leq n} \{ k - s(k) \}\]
\begin{codebox}
\zi \Comment setting Ex and Pr flags
\li \For $\forall k \in P$
	 \Do
\li		\If $k.local\_index < s^* \quad and \quad k.Ex \; != \;true$
\li		\Then $k.Pr=true$
			\End
\li		\If $k.local\_index < I$
\li		\Then $k.Ex=true$
			\End
		\End
\end{codebox}

\subsubsection{Results and conclusions}

It is obvious that the proposed algorithm generates access patterns which has lower S\_BW than a given bound. The input length multiplier \textbf{k} is a good parameter for measuring the solution quality. (k-1)*100\% of the vertices have to be reloaded from the main memory, but the processing still has 0\% cache-miss\footnote{assuming the handling of multiplicity problems}.\\

\begin{table}[h!]
\begin{center}
\caption{Results of AM1 bounded bandwidth optimization}
\label{AM1-bounded}
\setlength{\tabcolsep}{3pt}
\begin{scriptsize}
\begin{tabular}{|c|c|c|c|c|c|c|c|}
\hline
Case & AM1\_BW & S\_BW Bound & num. of parts & N & overall length & k & time(s)\\ \hline
3d\_075 & 391 &	392 &	 1  &	 3562 &	3562 &	1  &	0,255\\ \hline
3d\_075 & 391 &	380	&  4	&  3562	& 4288 &	1,203 &	0,392\\ \hline
%3d\_075 & 391 &	350	&  5	&  3562	& 4490 &	1,260	& 0,522\\ \hline
3d\_075 & 391 &	300	&  9	&  3562	& 4945 &	1,388	& 0,7\\ \hline
3d\_075 & 391 &	200	& 20	&  3562	& 5929 &	1,664	& 1,148\\ \hline \hline
%3d\_075 & 391 &	100	&109	&  3562	&10334 &	2,901	& 5,45\\ \hline
%3d\_075 & 391 &	50	&423	&  3562	&18479 &	5,187	& 30,725\\ \hline 
3d\_035 &1863 &	1864&	1	  & 33730	&33730 &	1     &	2,317\\ \hline
3d\_035 &1863 &	1800&	6	  & 33730	&38053 &	1,128	& 7,78\\ \hline
3d\_035 &1863 &	1500&	5	  & 33730	&38702 &	1,147	& 4,439\\ \hline
%3d\_035 &1863 &	1000&	19	& 33730	&43352 &	1,285	& 6,047\\ \hline
3d\_035 &1863 &	500	& 88	& 33730	&58171 &	1,724	& 36,095\\ \hline \hline
%3d\_035 &1863 &	250	&290	& 33730	&78082 &	2,314	& 213,219\\ \hline
3d\_015 &14930& 14931&	1	  &417573	&417573&	1	    & 70,108\\ \hline
3d\_015 &14930& 14000&	2	  &417573	&431081&	1,032	& 77,004\\ \hline
3d\_015 &14930& 10000&	2	  &417573	&427211&	1,023	& 71,278\\ \hline
3d\_015 &14930&	7500&	8	  &417573	&449441&	1,076	& 91,058\\ \hline
3d\_015 &14930&	5000&	34	&417573	&476170&	1,140	& 53,247\\ \hline
3d\_015 &14930&	2500&	130	&417573	&557190&	1,334	& 687,385\\ \hline
\multicolumn{8}{l}{AM1\_BW: the bandwidth provided by AM1 for the whole mesh}\\
\multicolumn{8}{l}{overall length: length of the generated access pattern}\\
\multicolumn{8}{l}{N: number of vertices, Algorithm tested on one core of an Intel P8400 processor}\\
\end{tabular}
\end{scriptsize}
\end{center}
\end{table}

Measurements on three meshes with different S\_BW bounds can be found on Table~\ref{AM1-bounded}. The results shows that the solution quality mainly depends on the distance of the S\_BW bound and the maximal\_degree of the mesh. This is a really good news, because maximal\_degree is around 20-30 for a typical 3D mesh, while the S\_BW bound is around 10-40k\footnote{the bound is depending on sizeof(node)} nowadays and increasing with each new generation of FPGA-s. The number of generated parts determine the time consumption of the proposed method, because in each restart of AM1 the algorithm calculates the pseudo diameter for the rest nodes. The results on 3d\_015 shows that we can go below 25\% of the original S\_BW with 15-30\% reload. This method gives the opportunity of deciding the size of the on-chip memory synthesized to the FPGA, so the designers can have more free area with sacrificing computational time.\\

\section{Arithmetic unit generation}

The VHDL representation of the arithmetic unit optimized for speed and area was automatically generated via the framework presented at \cite{Nemes2011}.
To reach high operating frequency the floating-point units shall be partitioned and a local control unit shall be assigned to every cluster.
The objective is to minimize the number of extra FIFOs required for data synchronization between the clusters while
guaranteeing that the signals of the control unit have tolerable fanout and do not decrease the operating frequency of the rest of the circuit.
In \cite{Nemes2011} it was demonstrated that the placement of a partition generated by traditional partitioniers which minimize the previous objective is very challenging and the maximum operating frequency of the slowest floating-point unit cannot be reached with the standard Xilinx place and route tools.
A new partitioning strategy was demonstrated which produces more cut nets than common strategies however the resulting partition can be easily mapped to the FPGA and the resulting circuit can operate near to the maximum operating frequency of the slowest floating-point units.

The main idea of the algorithm is to draw the graph into the plane before the partitioning starts. If a representation of the graph which minimizes the distance between the connected edges is given a simple greedy clustering algorithm can provide a partitioning without long interconnections between the clusters.

The input of the procedure is a text file containing the discretized state equations of the state variables.
In the first step the text file is parsed and a data-flow graph is created where every global variable and mathematical operator is represented with a vertex.
As there are enough resources on the FPGA and our goal is to reach the maximum operating frequency the data-flow graph is not optimized to reduce the number of vertices however the same effect can be reached manually if local variables are defined in the input file.

The next step of the initialization of the algorithm is to create a bipartite graph from the original data-flow graph. As every vertex representing an operator is associated with a delay time a bipartite graph can be easily created via
a breadth-first search which visits every vertex of the graph and computes the level of the vertex and the time required for the input to reach the given vertex. If the levels of its ancestors are different the algorithm inserts the proper number of extra vertices (delays) after the problematic ancestor. In physical implementation these delays will be shift registers which hold the data for the proper number of clock cycles and the computed levels determines the vertical coordinates of the vertices.

The proposed algorithm consists of two greedy phases. The first phase collects global information about the graph
by positioning the vertices horizontally, while clusters are created in the second phase using the information encoded in the spatial position of the vertices.

The first greedy phase tries to minimize the distance between the connected vertices described by the following equation:
$$
distance(A, B) := \left \{ \begin{array}{ll}
(x_A-x_B)^2&\mbox{if A and B are connected}\\
0&\mbox{otherwise}\\
\end{array} \right.
$$
where $x_A$ and $x_B$ are the horizontal coordinates of vertex $A$ and $B$ respectively.
A simple swap based algorithm like Kernighan-Lin \cite{KL} have been designed which minimize this objective for all vertices together.
This algorithm can be easily trapped in the local minimum therefore the initial placement of the vertices is
critical.
In our experiments the fast Barycenter heuristic \cite{Sugiyama} was used to create an initial solution however more complex techniques can be used as well. The Barycenter heuristic was initialized with several random placements and the best solution was seeded to the swap based iterative algorithm.

The second phase is another simple greedy algorithm which search the result of the first phase for rectangular clusters. Height of the rectangular domains can be chosen arbitrary however in our examples it is set to two. The algorithm is started from the top left corner and the largest possible rectangular cluster is searched which still meets the I/O constraint. Next the algorithm moves right and search for the largest possible rectangular cluster from the rest of the unclustered vertices. If there are no more unclustered vertices in the selected layers the algorithm moves down and continues with the lower layers.

Position of specific parts of the circuit can be constrained by the Xilinx PlanAhead software.
It enables the user to create rectangular placement constraints also called pblocks.
The Xilinx P\&R tools are likely to disperse the registers of the FIFOs which can limit the operating frequency of the circuit therefore separate pblocks were created for each FIFO buffer.
As the Xilinx P\&R tools were able to place and route the floating-point units inside the clusters no pblock were created for the floating-point units just for the clusters. 
The pblocks was placed manually using the positions of the placed vertices however in a future work this can be fully automatized if the area requirements of the clusters are also calculated. 

\section{Case study: Finite volume solver for the Euler equations}

In this section we describe the computational modelling of compressible fluid and gas flows by some of the basic tools available in the field of Computational Fluid Dynamics (CFD). The art of CFD is heavily exploiting the enormous processing power available by recent computer technology. Indeed, its central concept is the approximation of the continuous model problem by a discrete one, requiring the processing and the manipulation of a huge amount of data.

\subsection{Fluid Flows}\label{sect_fluidflow}

A wide range of industrial processes and scientific phenomena involve gas or fluids flows over complex obstacles, e.g. air flow around vehicles and buildings, the flow of water in the oceans or liquid in BioMEMS. In such applications the temporal evolution of non-ideal, compressible fluids is quite often modelled by the system of Navier-Stokes equations. The model is based on the fundamental laws of mass-, momentum- and energy conservation, including the dissipative effects of viscosity, diffusion and heat conduction. By neglecting all non-ideal processes and assuming adiabatic variations, we obtain the Euler equations \cite{Anderson1995, Chung2002}, describing the dynamics of dissipation-free, inviscid, compressible fluids. They are a coupled set of nonlinear hyperbolic partial differential equations, in conservative form expressed as
\begin{subequations}
\begin{equation}\label{eqContinuity}
%\small
\frac{{\partial \rho }}
{{\partial t}} + \nabla  \cdot \left( {\rho {\mathbf{v}}} \right) = 0
\end{equation}
\begin{equation}\label{eqMomentum}
%\small
\frac{{\partial \left( {\rho {\mathbf{v}}} \right)}}
{{\partial t}} + \nabla  \cdot \left( {\rho {\mathbf{vv}} + \hat Ip} \right) = 0
\end{equation}
\begin{equation}\label{eqEnergy}
%\small
\frac{{\partial E}}
{{\partial t}} + \nabla  \cdot \left( {(E + p){\mathbf{v}}} \right) = 0
\end{equation}
where $t$ denotes time, $\nabla$ is the Nabla operator, $\rho$ is the density, $u$, $v$ are the $x$- and $y$-component
of the velocity vector \textbf{v}, respectively, $p$ is the thermal pressure of the fluid, 
$\hat I$ is the identity matrix, and $E$ is the total energy density defined by
\begin{equation}\label{eqEnergyDensity}
%\small
E = \frac{p}
{{\gamma  - 1}} + \frac{1}
{2}\rho {\mathbf{v}} \cdot {\mathbf{v}}.
\end{equation}
\end{subequations}
In equation (\ref{eqEnergyDensity}) the value of the ratio of specific heats is taken to be $\gamma=1.4$. For later use we introduce the conservative state vector $\mathbf{U}=[ \rho, \rho u, \rho v, E]^T$ , the set of primitive variables $\mathbf{P}=[ \rho, u, v, p]^T$  and the speed of sound $c = \sqrt {\gamma p/\rho }$. It is also convenient to merge (\ref{eqContinuity}), (\ref{eqMomentum}) and (\ref{eqEnergy}) into hyperbolic conservation law form in terms of U and the flux tensor
\begin{equation}\label{eqFlux}
%\small
{\mathbf{F}} = \left( {\begin{array}{*{20}c}
   {\rho {\mathbf{v}}}  \\
   {\rho {\mathbf{vv}} + Ip}  \\
   {(E + p){\mathbf{v}}}  \\
 \end{array} } \right)
\end{equation}
as:
\begin{equation}\label{eqConsLaw}
%\small
\frac{{\partial U}}
{{\partial t}} + \nabla  \cdot {\mathbf{F}} = 0.
\end{equation}

\subsection{Discretization of the governing equations}\label{sect_discr}

Logically structured arrangement of data is a convenient choice for the efficient operation of the FPGA based implementations \cite{KOCSARDI2008}. However, structured data representation is not flexible for the spatial discretization of complex geometries. As one of the main innovative contributions of this paper, here we consider an unstructured, cell-centered representation of physical quantities. In the following paragraphs we describe the mesh geometry, the governing equations, and the main features of the numerical algorithm.

\subsubsection{The geometry of the mesh}\label{Section3.1}

The computational domain $\Omega$ is composed of non-overlapping triangles. The $i$-th face of triangle $\cal{T}$ is labelled by $f_i$. The normal vector of $f_i$ pointing outward $\cal{T}$ that is scaled by the length of the face is ${\bf n_i}$.
The volume of triangle $\cal{T}$ is $V_{\cal{T}}$. Following the finite volume methodology, all components of the volume averaged quantities are stored at the mass center of the triangles.

\subsubsection{The Discretization Scheme}\label{Section3.2}

\begin{figure}
\begin{center}
\epsfig{width=8cm,figure=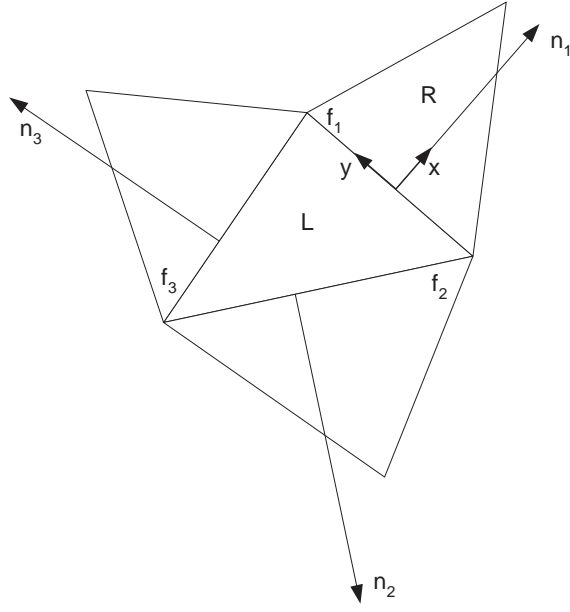}
\end{center}
\vspace{-0.2cm}
\caption{Interface with the normal vector and the cells required in the computation} \label{FigInterface}
\vspace{0.2cm}
\end{figure}

%In this paper we consider a second-order algorithm both in space and time. 
Application of the cell centered finite volume discretization method leads to the following semi-discrete form of governing equations (\ref{eqConsLaw})
\begin{equation}\label{eqSemiDiscrete}
\frac{{dU_{\cal{T}} }}
{{dt}} =  - \frac{1}
{{V_{\cal{T}} }}\sum\limits_f {{\mathbf{F}}_f  \cdot {\mathbf{n}}_f },
\end{equation}
where the summation is meant for all three faces of cell ${\cal{T}}$, and
\textbf{F}$_f$ is the flux tensor evaluated at face $f$.
Let us consider face $f$ in a coordinate frame attached to the face, such, that its $x$-axes is normal to $f$ (see Fig.~\ref{FigInterface}). 
Face $f$ separates triangle L (left) and triangle R (right).
In this case the $\textbf{F}_f  \cdot \textbf{n}_f$ scalar product equals to the x-component of \textbf{F} (i.e. F$_x$) multiplied by the area of the face. In order to stabilize the solution procedure, artificial dissipation has to be introduced into the scheme. According to the standard procedure, this is achieved by replacing the physical flux tensor by the numerical flux function F$^N$ containing the dissipative stabilization term. A finite volume scheme is characterized by the evaluation of F$^N$, which is the function of both U$_L$ and U$_R$. In this paper we employ the simple and robust Lax-Friedrichs numerical flux function defined as
\begin{equation}\label{eqCompFlux}
F^N  = \frac{{F_L  + F_R }}
{2} - \left( {\left| {\bar u} \right| + \bar c} \right)\frac{{U_R  - U_L }}{2}.
\end{equation}
In the last equation F$_L$=F$_x$(U$_L$) and F$_R$=F$_x$(U$_R$) and notations $\left| {\bar u} \right| $ and  $\left| {\bar c} \right|
$ represent the average value of the $u$ velocity component and the speed of sound at an interface, respectively.  The temporal derivative is dicretized by the first-order forward Euler method:
\begin{equation}\label{eqFWEuler}
\frac{{dU_{\cal{T}} }}
{{dt}} = \frac{{U_{\cal{T}}^{n + 1}  - U_{\cal{T}}^n }}
{{\Delta t}},
\end{equation}
where $U_{\cal{T}}^n$ is the known value of the state vector at time level~$n$, $U_{\cal{T}}^{n+1}$ is the unknown value of the state vector at time level~$n+1$, and $\Delta t$ is the time step.

By working out the algebra described so far, leads to the discrete form of the governing equations to compute the numerical flux term F and the dissipation term D. 
\begin{equation}
U_{\cal{T}}^{n+1} = U_{\cal{T}}^{n} -\frac{\Delta t}{V_{\cal{T}}}
\sum\limits_{f} \hat{\cal R}_{\bf n_f} F_f |{\bf n_f}|,
\end{equation}

\noindent
where $\hat{\cal R}_{\bf n_f}$ is the rotation tensor describing the transformation from the normal-parallel coordinate frame of face $f$ to the $x-y$ frame. Quantity $F_f$ is defined in a coordinate frame attached to face $f$, with such an orientation that state left is identical to the state of the update triangle ${\cal T}$ while state right corresponds to the state of the triangle situated at the opposite side of the face. With these conventions, the normal component of the numerical flux function is given by

\begin{subequations}
\begin{align}
F_f^{\rho} &= \frac{{\rho_L u_L  + \rho_R u_R }}{2} \label{eqFi}
+\left(\left| \bar u \right| + \bar c \right)\frac{\rho _R  - \rho _L}{2}\\
F_f^{\rho u} &= \frac{{\left( {\rho_L u_L^2  + p_L} \right) + \left( {\rho_R u_R^2  + p_R} \right) }}{2}
+\left(\left| \bar u \right| + \bar c \right)\frac{\rho_R u_R  - \rho_L u_L}{2}\\
F_f^{\rho v} &= \frac{{\rho_L u_Lv_L  + \rho_R u_Rv_R }}{2}
+\left(\left| \bar u \right| + \bar c \right)\frac{\rho_R v_R  - \rho_L v_L}{2}  \\
F_f^{E} &= \frac{{\left({E_L + p_L} \right)u_L  + \left( {E_R + p_R} \right)u_R }}{2} 
+ \left(\left| \bar u \right| + \bar c \right)\frac{E_R  - E_L}{2}
\end{align}
\end{subequations}

\begin{figure}
\begin{center}
\epsfig{width=15cm,figure=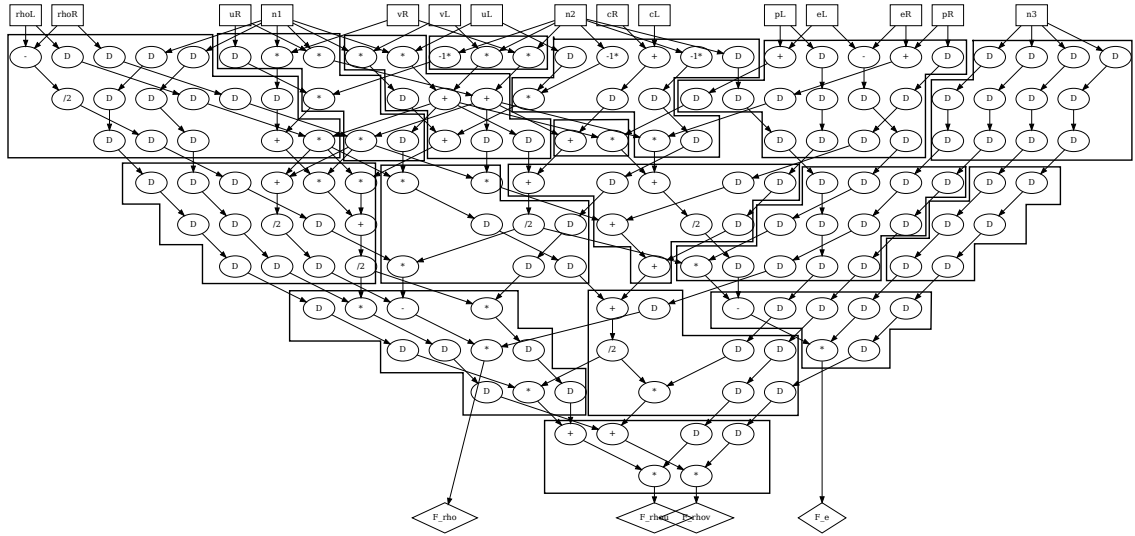}
\end{center}
\caption{The partitioned arithmetic unit}\label{FigCFDPart}
\end{figure}

\section{Results, performance}
The generated arithmetic unit will be implemented on our AlphaData ADM-XRC-6T1 reconfigurable development system \cite{AlphaData} equipped with a Xilinx Virtex-6 XC6VSX475T FPGA \cite{Xilinx} and 2Gbyte on-board DRAM in four 32bit wide banks running on 8000MHz. 

In our test case the cell centered approach is used therefore each triangle is represented by a node and each node has 3 neighbors except at the boundaries where ghost nodes are used to implement inflow, outflow or wall boundary conditions. For simplicity size and normal vector of all faces are precomputed which results in higher memory bandwidth requirement, but the arithmetic unit will be simpler. The architecture is implemented using both single and double precision floating point numbers. The node data structure contains four time dependent variables $[\rho, \rho u, \rho v, E]$ and the area of the triangle as a constant additionally pressure $p$ and local speed of sound $c$ which is computed on the FPGA for each node also should be stored. Therefore width of data bus of the Memory unit is 28 and 56 byte in the single and double precision cases. To provide this wide memory bus 14 BRAMs should be used in a $32\times512$ configuration. When all BRAMs are allocated for the Memory unit 38,912 nodes can be stored on the FPGA which is usually more than enough for practical 2D meshes.

Depending on the width of the values the input data bus is 20 or 80 byte wide while 16 or 32 byte wide results should be loaded and saved in every 3 clock cycles. Additionally 3 face descriptors should be loaded which contains 9 floating point values and $3\times2$ byte wide indices altogether. Therefore 10.3Gbyte/s or 19.7Gbyte/s memory bandwidth is required to feed the arithmetic unit with valid data in each clock cycle. In the double precision case this bandwidth cannot be provided on our prototyping board and the system will be memory bandwidth limited. It should be noted that computing face normals on the FPGA using the coordinates of the vertices results in about $30\%$ smaller memory bandwidth requirement at a price of more complicated on-chip memory structure and arithmetic unit. Results from the output of one processor can be feed directly to another processor linearly increasing the performance of the architecture without additional bandwidth requirement.

\subsection{Test setup}
To show the efficiency of our solution a complex test case was used, in which a Mach 3 flow over a forward facing step was computed. The simulated region is a two dimensional cut of a pipe which has closed at the upper and lower boundaries, while the left and right boundaries are open. The direction of the flow is from left to right and the speed of the flow at the left boundary is 3-time the speed of sound constantly. The solution contains shock waves reflecting from the closed boundaries. 

\begin{figure}
\begin{center}
\epsfig{width=15cm,figure=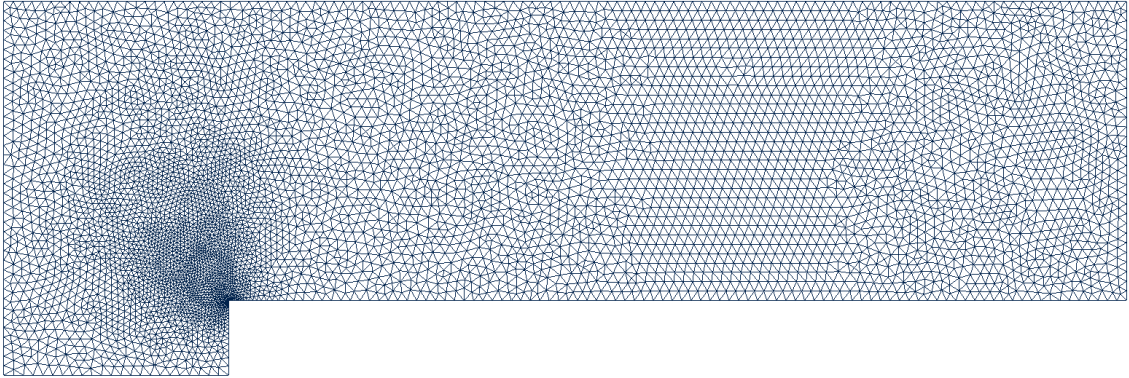}
\end{center}
\caption{Coarse resolution mesh for the forward facing step test case}\label{FigGridStepcl40}
\end{figure}

Unstructured mesh for the domain are generated by the freely available Gmsh mesh generator \cite{Geuzaine2009}. Several meshes are generated using different characteristic length between $1/30.0-1/200.0$ while characteristic length of the elements near the corner of the step is divided by $12.5$. The number of triangles in the resulting meshes is ranging in the 7,063-394,277 interval. An example mesh generated by using $1/40.0$ characteristic length and containing 12,297 triangles is shown in Figure~\ref{FigGridStepcl40}.

\subsection{Performance comparison}
During a comparison various mesh sizes are used with 7,063 to 394,277 triangles. Performance of our architecture is estimated using the result of static timing analysis which is indicated 390MHz operating frequency in the double precision case. Three clock cycles are required to update the state of one triangle therefore performance of one processor is 130million triangle update/s. Computation of one triangle requires 213 floating point operations therefore performance of our architecture is 27.69GFLOPs. On the Virtex-6 XC6VSX475T FPGA three arithmetic units can be implemented resulting in 83.07GFLOPs cumulative computing performance.

\begin{figure}
\begin{center}
\epsfig{height=8cm,angle=-90,figure=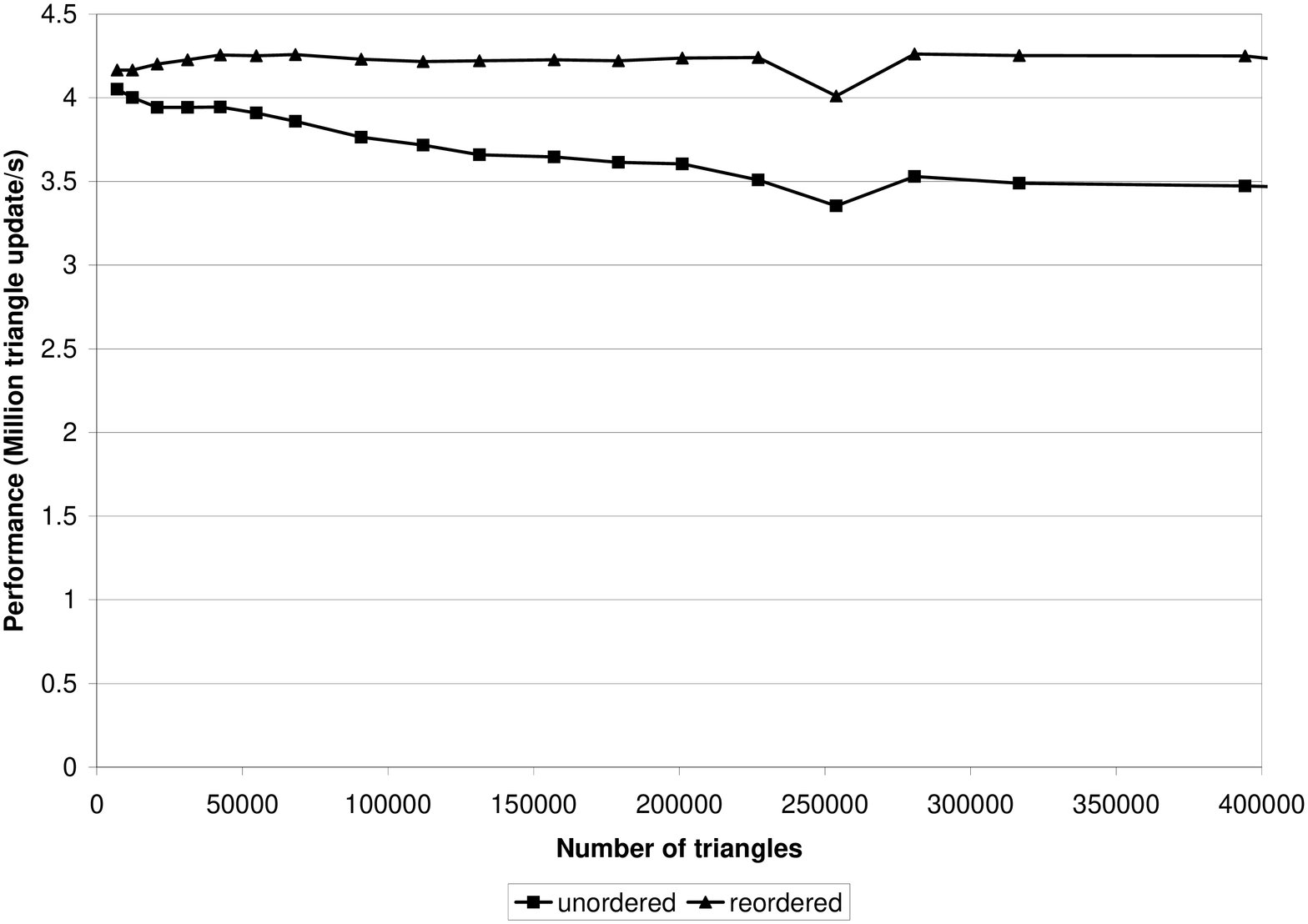}
\end{center}
\caption{Measured performance of Intel Xeon E5620 microprocessor}\label{FigCPUPerfUnstr}
\end{figure}

Performance of our architecture is compared to a high performance Intel Xeon E5620 microprocessor running on 2.4GHz clock frequency. On the microprocessor a single core is used and the simulation is carried out with and without renumbering the nodes. Performance of the simulations is shown in Figure~\ref{FigCPUPerfUnstr}. As we expected without renumbering the performance of the simulation is steadily decreasing as the mesh size is increased while renumbering the nodes improved the performance of the microprocessor by $15\%$ to 4.22million triangle update/s or equivalently 898.86MFLOPs.

Comparison of the performance of the two architectures show that a single processor implemented on the FPGA can outperform the Intel Xeon processor computing 30 times faster. By connecting three processors operating in parallel on a single FPGA achieve linear speedup and provide 90 times more computing power than a microprocessor.
%During the comparison only one CPU core is used. Ideally 8 times more computing power can be achieved when all the four cores and SSE vector instructions are used.

\begin{figure}
\begin{center}
\epsfig{width=15cm,figure=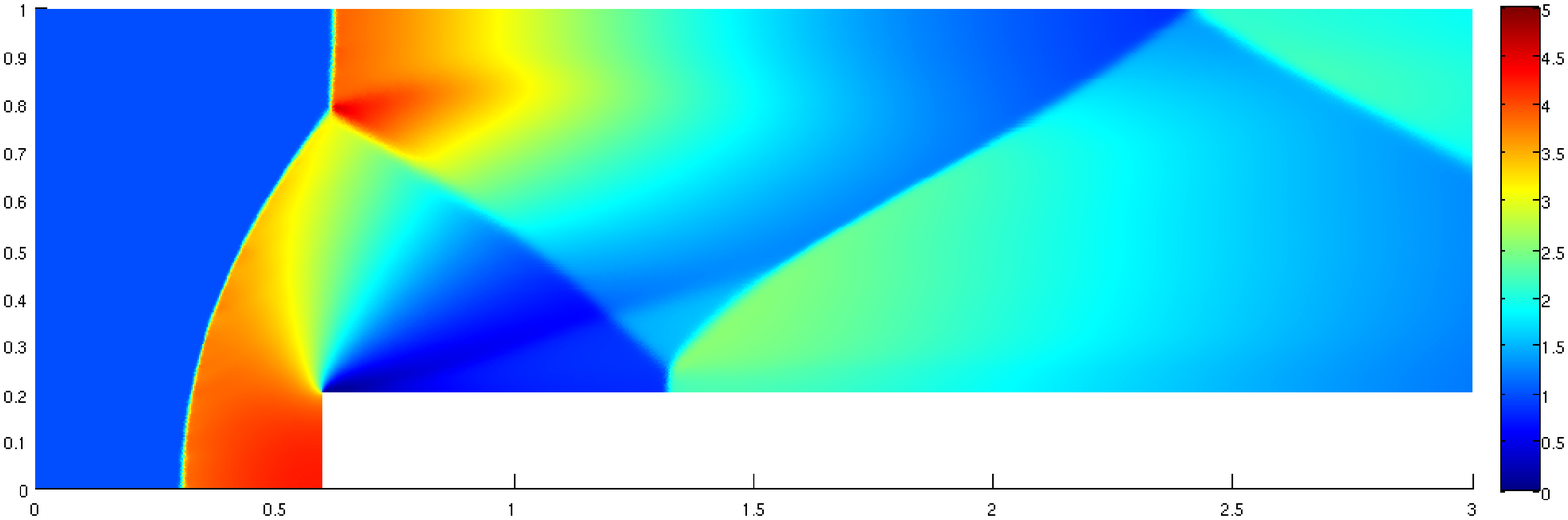}
\end{center}
\caption{First-order solution of the Mach 3 flow after 4s simulation time on a 394,277 triangle mesh} \label{FigMach31stUnstr}
\end{figure}

Result of the computation on the largest simulated grid after 4s of simulation time is shown in Figure~\ref{FigMach31stUnstr}. Reference solution for the previous problem computed by the more accurate residual distribution upwind scheme can be found in \cite{Csik2002}.

\section{Conclusions}
A framework for accelerating the solution of PDEs using explicit unstructured finite volume discretization is presented. Irregular memory access patterns can be eliminated by using the proposed memory structure which results in higher available memory bandwidth and full utilization of the arithmetic unit. Efficient use of the on-chip memory is provided by a new node reordering algorithm which can be extended to generate fixed bandwidth partitions. The new algorithm is comparable to the well known GPS algorithm in both runtime and quality of the results.

Generation of the application specific arithmetic unit is described by using a complex numerical problem solving the Euler equations. The discretized state equations are automatically translated to a synthesizable VHDL description using Xilinx floating-point IP cores. Performance of the arithmetic unit is improved by using partitioning and local control units. Nodes of the arithmetic unit is rearranged to minimize edge crossing which helped placement of the partitions and  improved clock frequency of the design.

Performance comparison of the architecture using a single processor running on 390MHz showed that 30 times speedup can be achieved compared to a high performance microprocessor core. Computing performance can be further improved by implementing three processors on one FPGA reaching 90 times speedup.

Currently size of the mesh is limited by the bandwidth of its adjacency matrix which must be smaller than 40,000. The architecture should be improved to efficiently handle multiple partitions and extended to use multiple FPGAs during computation.

\section*{Acknowledgments}
This research project supported by the János Bolyai Research Scholarship of the Hungarian Academy of
Sciences, TAMOP-4.2.1./B-10, TAMOP-4.2.1./B-11, OTKA Grant No. K84267 and in part by OTKA Grant
No. K68322.
%This research project supported  by  the  J\'{a}nos  Bolyai  Research Scholarship of the Hungarian Academy of Sciences and in part by OTKA Grant No. K68322.%CSak egy OTKA???? !!!!!!!!!!!

\bibliographystyle{unsrt}
\bibliography{references}
\end{document}